\documentclass{template/aa}  

\usepackage{graphicx}
\graphicspath{{figures/}}
\usepackage{txfonts}
\usepackage{lipsum}
\usepackage{subcaption}  
\usepackage{placeins} 
\usepackage{xcolor}
\usepackage{hyperref}

\DeclareFontFamily{U}{wncy}{}
\DeclareFontShape{U}{wncy}{m}{n}{<->wncyr10}{}
\DeclareSymbolFont{mcy}{U}{wncy}{m}{n}
\DeclareMathSymbol{\sha}{\mathord}{mcy}{"58} 

\newcommand{\conv}{\ast}
\newcommand{\phires}{\varphi_\text{res}}
\newcommand{\phidm}{\varphi_\text{DM}}
\newcommand{\phiatm}{\varphi_\text{atm}}
\newcommand{\psd}[1]{\langle|#1|^2\rangle}
\newcommand{\sensi}{\mathcal{M}}
\newcommand{\IFFT}[1]{\mathcal{F}^{-1}\left\{#1\right\}}

\newcommand{\aopera}{\texttt{AOPERA} }
\newcommand{\oopao}{\texttt{OOPAO} }
\newcommand{\tiptop}{\texttt{TIPTOP} }

\newcommand{\rfe}[1]{#1}

\begin{document}

   \title{Modelling Fourier filtering wavefront sensors for PSD-based methods: the AOPERA tool}

   \subtitle{}

   \author{Romain JL Fétick\inst{1,2}
        \and Vincent Chambouleyron\inst{2}
        \and Lisa-Marie Mazzolo\inst{1,2}
        \and Christophe Vérinaud\inst{3}
        \and Arnaud Striffling\inst{4}
        \and Cédric Taissir Héritier-Salama\inst{1,2}
        \and Benoit Neichel\inst{2}
        \and Jean-François Sauvage\inst{1,2}
        \and Thierry Fusco\inst{1,2}
        }

   \institute{DOTA, ONERA, 13330, Salon-de-Provence, France\\
             \email{romain.fetick@onera.fr}
            \and Aix Marseille University, CNRS, CNES, LAM, Marseille, France
            \and European Southern Observatory (ESO), Karl Schwarzschild-str 2, D-85748 Garching-bei-Muenchen Germany
            \and Aix Marseille Univ, CNRS, Pytheas, OHP, Observatoire de Haute-Provence, France
            \\ }

   \date{Received September 30, 20XX}

  \abstract
   {Compensation of the atmospheric turbulence thanks to adaptive optics (AO) has now become commonly used for VLT or ELT class telescopes in order to retrieve a resolution close to their diffraction limit.}
   {Following the increasing trend of AO system, there is also a stronger necessity for simulations in order to understand and predict their performance facing different observing conditions, that are the evolution of the atmospheric turbulence or the diversity of AO guide source. The so-called PSD based methods are well adapted to the demand thanks to their simplicity and speed. Moreover, they provide a comprehensive breakdown error budget and impact on focal plane, that is of high interest especially in the case of extreme adaptive optics. Their drawback is the challenge to describe non-linear wavefront sensors (WFS) such as the pyramid WFS or the Zernike WFS. There is a necessity to develop fast yet accurate methods to describe the behaviour of AO systems including sensitive WFS.} 
   {The convolutive formalism is a powerful mathematical tool to describe the sensitivity of Fourier filtering wavefront sensors. We thus develop a method to compute the AO system response (electromagnetic phase power spectral density, and point spread function) based on the accuracy of the convolutive formalism. Moreover, our method allows to include the non-linear behaviour of the wavefront sensor within PSD-based numerical tools. Mathematical formalism to describe the FF-WFS sensitivity combined to non-linearity management greatly improve the accuracy of description of AO systems through numerical simulations.}
   {After a mathematical description of the method, its numerical implementation is compared with end-to-end simulations using the \oopao tool. Indeed end-to-end simulations are reproducing the response of AO systems with high fidelity, especially regarding WFS sensitivity and non-linearity. Similarity of results for the fitting error, temporal error and noise error proves the validity of our method in managing Fourier filtering WFS sensitivity and non-linearity. It also cross-validates both numerical implementations of our method and \oopao as a justification for any future usage of both tools.}
   {Development of a PSD-based method accurately describing Fourier filtering WFS and justification of its validity is a critical technique for understanding the behaviour of AO systems during their design or operation phases. The model and its numerical implementation have been applied to the PROVIDENCE $2.5$ m telescope to be built in France by the end of 2028, and could be used for ELT instruments which include a pyramid AO system.}


   \maketitle
   
   \nolinenumbers

\section{Introduction}

Motivated by the successful pioneering work of the past decades and fast technological developments, adaptive optics (AO) has become a commonly used technique for nowadays astronomical observations. Indeed multiple astronomical objectives require high angular resolution for precise astrometry \citep{genzel2010}, high contrast imaging in the vicinity of a host star \citep{beuzit2008}, or spatial filtering with injection into single mode fibers to feed high resolution spectrographs \citep{delorme2021,vigan2024,carlotti2022,lovis2022}. AO systems thus flourish to equip current very large telescopes \citep{sauvage2016,stuik2006,vandam2004,pinna2016} and will become the baseline of any instruments on the future extremely large telescopes \citep{brandl2016,diolaiti2016,davies2016,ciliegi2022,bond2022}. AO systems are also important for metric size telescopes such as PROVIDENCE \citep{petit2024}. However, performance of AO systems are highly dependent of the observing conditions, namely the turbulence and the source characteristics (magnitude and possible spatial extension). Understanding the response of the AO system facing highly variable conditions is critical for AO design, exposure time calculators (ETC), prediction of quality for the scientific data, or image post-processing.\\

A precise description of the AO systems is thus required, relying on numerical simulations from the system design to commissioning and eventually operations. Exquisite precision can be obtained with the so-called end-to-end simulators, consisting in generating random phase screens and propagating them through the optical path of the instrument. The diversity of simulators includes YAO \citep{rigaut2013}, COMPASS \citep{gratadour2014}, HCIPY \citep{por2018}, \rfe{SOAPY \citep{reeves2016}, DASP \citep{basden2018}}, PASSATA \citep{agapito2016}, SPECULA \citep{rossi2026}, OOMAO \citep{conan2014oomao} and OOPAO \citep{heritier2023oopao}. These end-to-end techniques can model fine effects such as wavefront sensor (WFS) non-linearity, non-linear controllers, transient effects or AO loop instabilities. The drawback is an extensive numerical computation cost, that increases with the system complexity, especially with the loop frequency and the number of actuators. Indeed, for a typical system running at $1$ kHz, a pupil of $10$ m class telescope and a wind speed of $10$ m/s, it thus requires $10\,000$ iterations to slide the turbulence over ten times the pupil diameter in order to obtain statistical representation of the low orders. Moreover, these simulations encompass at once all error terms of the AO system such as the fitting error, temporal error, or aliasing error. Splitting the error budget of these end-to-end simulations is made possible by extra tools such as ROKET \citep{ferreira2018}. As stated by the authors, ROKET is able to provide a scalar error budget of end-to-end simulations using linear WFS. However, to the best of our knowledge, there is yet no development for non-linear wavefront sensors such as the pyramid WFS. Finally, there is no splitting of the phase spatial PSD by error term, only the covariance of the error terms is available.\\

A complementary method to the end-to-end simulations considers the phase statistics, namely its power spectral density (PSD), instead of random samples of phase screens. These tools are often referred as PSD-based \rfe{\citep{males2018,juvenal2018,neichel2021tiptop}}. Being fast, the PSD-based method is adapted to ETC or parametric study for AO systems. The PSD method is numerically frugal as it does not scale up with the loop frequency. Moreover the error budget can be split into different error terms, allowing to focus on the major AO error terms to be tackled, and thus offering the possibility to optimize an AO system during its design phase. Such tools are even adapted to machine learning techniques for point-spread-function (PSF) prediction \citep{kuznetsov2023}.\\

The major drawback of PSD-based method is that they assume a linear filtering of the AO system on the turbulence and noise PSDs. The advent of Fourier-filtering \citep{fauvarque2017phd} wavefront sensors (FF-WFS), such as the pyramid WFS or the Zernike WFS, challenges the PSD based methods. Indeed these FF-WFS have been developed for their high sensitivity with respect to the incoming phase, thus reducing the propagation of noise. It allows wider sky coverage for AO systems, enables fast correction in the case of extreme adaptive optics \citep{guyon2011,males2016} or double stage adaptive optics \citep{cerpa2022}. However this high sensitivity is obtained at the cost of a non-linear response of the WFS with respect to the input phase. We developed a method based on the convolutive formalism \citep{Fauvarque2016formalism} for FF-WFS, allowing for a precise estimation of the FF-WFS sensitivity. We also take into account the non-linear behaviour of the FF-WFS, reducing its sensitivity when facing strong phase residuals, leading to the so-called optical gains (OG) defined as the ratio of sensitivity between calibration and on-sky operations. Taking into account the non-linear behaviour of FF-WFS is critical for AO design, since ignoring it would always favor the most sensitive WFS independently of the observing conditions. Combining both accurate sensitivity estimation and non-linearity management greatly improves the fidelity of PSD based method to describe FF-WFS systems.\\

Section~\ref{sec:method} describes the mathematical formalism developed to take into account spatial and temporal filtering of the AO system, the usage of the convolutive formalism and the method used to take into count FF-WFS non-linearity. Section~\ref{sec:tiptop} locates our work with respect to state-of-the-art tools. Section~\ref{sec:results} validates our method against precise end-to-end simulations (using the OOPAO tool) that is an important justification of the non-linear PSD based method. Finally, section~\ref{sec:applications} describe direct applications of our tool on extended source observation, that is a challenging topic for FF-WFS.

\section{Method}
\label{sec:method}

This section recalls the usual method to derive a PSD budget for an adaptive optics system \citep{rigaut1998,jolissaint2010,neichel2021tiptop} and the three major usual hypothesis that are error terms independence, frozen flow of the turbulent layers and temporally white noise. Then the optical gain effects are included into the equations thanks to the convolutive formalism, allowing for non-linear behaviour of the WFS.

\subsection{Residual phase on the WFS}

\begin{figure}[h]
    \centering
    \includegraphics[width=0.9\linewidth]{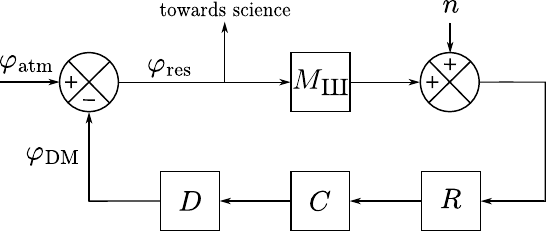}
    \caption{Schematics of the adaptive optics loop with the notations used in this paper: sensitivity $\sensi_\sqcap(\vec{f})$, reconstructor $\mathcal{R}(\vec{f})$, temporal aspect $\mathcal{C}(\nu)$, deformable mirror influence $\mathcal{D}(\vec{f})$.}
    \label{fig:schema-oa}
\end{figure}

Adaptive optics (AO) is an automatic system working in closed loop on the electromagnetic phase in order to maintain it towards a given reference phase. The AO system acts both as a spatial and temporal filter on the phase, so we choose to remember these dependencies in the following equations by noting $\vec{f}=(f_x,f_y)$ the spatial frequencies and $\nu$ the temporal frequencies. The residual phase $\phires$ seen on the wavefront sensor is the difference between the incoming atmospheric phase $\phiatm$ \rfe{(based on the Von K\'arm\'an turbulence model for our developments, but any model may be used)} and the deformable mirror correction phase $\phidm$ (see Figure\ref{fig:schema-oa}), such as:
\begin{equation}
    \phires(\vec{f},\nu) = \phiatm(\vec{f},\nu) - \phidm(\vec{f},\nu)
\end{equation}
The DM phase itself is due to both contributions of the measurement of the phase residuals on the WFS and the noise on the WFS camera, such as:
\begin{equation}
    \phidm(\vec{f},\nu) = \mathcal{D}(\vec{f}) \, \mathcal{C}(\nu) \, \mathcal{R}(\vec{f}) \left[ n(\vec{f},\nu) + \sha(\sensi\left(\vec{f})\,\phires(\vec{f},\nu)\right) \right]
\end{equation}
with $n$ the noise, $\mathcal{D}$ the DM correctable frequencies, $\mathcal{C}$ the temporal aspect (controller, WFS frame integration, loop delay, DM dynamics), $\mathcal{R}$ the spatial reconstructor, $\sensi$ the WFS measurement and $\sha$ the 2-dimensional discretization operator due to the WFS sampling. The latter writes:
\begin{equation}
\label{eq:aliasing}
\begin{split}
    \sha(\sensi\left(\vec{f})\,\phires(\vec{f},\nu)\right) = & \sum_{\vec{k}\in\mathbb{Z}^2} \sensi(\vec{f}+\vec{k}/p) \, \phires(\vec{f}+\vec{k}/p,\nu) \\
    & \,\,\,\,\,\, \times \text{sinc}(pf_x+k_x) \, \text{sinc}(pf_y+k_y)
\end{split}
\end{equation}
with $p$ the WFS pixel size in the telescope pupil plane. The index $\vec{k}=(0,0)$ corresponds to the direct measurement of the spatial frequencies, whereas all other indices correspond to the aliasing of higher spatial frequencies. For the sake of simplicity, let's define $\sensi_\sqcap$ as the continuous measurement $\sensi$ damped by the sinc functions from the WFS pixel spatial extension:
\begin{equation}
    \sensi_\sqcap (\vec{f}) \triangleq  \sensi(\vec{f}) \, \text{sinc}(pf_x) \, \text{sinc}(pf_y)
\end{equation}
and let's define $\Omega(\vec{f},\nu)$ the open loop transfer function:
\begin{equation}
    \Omega(\vec{f},\nu) \triangleq \mathcal{D}(\vec{f}) \, \mathcal{C}(\nu) \, \mathcal{R}(\vec{f}) \, \sensi_\sqcap(\vec{f})
\end{equation}
Splitting the index at $\vec{k}=(0,0)$ from the others ones within the discretized measurement equation, one finds the residual phase on the WFS as the sum of the three terms:
\begin{equation}
\label{eq:wfs-phase-error}
    \begin{split}
        \phires(\vec{f},\nu) = & \frac{1}{1+\Omega(\vec{f},\nu)} \phiatm(\vec{f},\nu) - \frac{\mathcal{D}(\vec{f}) \, \mathcal{C}(\nu) \, \mathcal{R}(\vec{f})}{1+\Omega(\vec{f},\nu)} n(\vec{f},\nu) \\
        & - \frac{\mathcal{D}(\vec{f}) \, \mathcal{C}(\nu) \, \mathcal{R}(\vec{f})}{1+\Omega(\vec{f},\nu)} \sum_{\vec{k}\in\mathbb{Z}^2\backslash (0,0)} \sensi_\sqcap(\vec{f}+\vec{k}/p) \, \phires(\vec{f}+\vec{k}/p,\nu)
    \end{split}
\end{equation}
with the sum running on $\mathbb{Z}^2\backslash (0,0)$ only, since the component $\vec{k}=(0,0)$ has been split apart and appears implicitly within the denominator $1+\Omega(\vec{f},\nu)$.\\

The first term of the residual phase equation~\ref{eq:wfs-phase-error} acts as a combined spatial and temporal filter on the turbulence, due to the term $\Omega(\vec{f},\nu)$, it thus encompasses both fitting error and temporal error. This combined term is called spatio-temporal error throughout this article. The second term of the equation~\ref{eq:wfs-phase-error} is a filter on the WFS noise (photon noise, readout noise, ...) and the last term is the aliasing of the high spatial frequencies onto the measurement.

\subsection{Hypothesis for PSD computation}
\label{sec:hypothesis}

\subsubsection{Non-correlation hypothesis}

The spatial PSD of the phase, noted $W(\vec{f})$, is computed by integration over the temporal frequencies:
\begin{equation}
    W(\vec{f}) = \int_{-\infty}^{+\infty} \psd{\varphi(\vec{f},\nu)} \, d\nu
\end{equation}
The PSDs for the three terms of equation~\ref{eq:wfs-phase-error} are considered statistically uncorrelated. This is valid since $\phires(\vec{f}+\vec{k}/p,\nu)=\phiatm(\vec{f}+\vec{k}/p,\nu)$ with $\vec{k}\neq (0,0)$ that is uncorrelated with $\phiatm(\vec{f},\nu)$. The PSDs are consequently summed to obtain the total residual PSD:
\begin{equation}
    W_\text{res}(\vec{f}) = W_\text{res,spatio-temporal}(\vec{f}) + W_\text{res,noise}(\vec{f}) + W_\text{res,aliasing}(\vec{f})
\end{equation}

\subsubsection{Temporally white noise hypothesis}

In this article, we consider that the measurement noise is white in the temporal direction, that is a valid assumption for photon noise and readout noise. The WFS signal is temporally sampled at the frequency $F$, let us call $\sigma^2_n$ the noise variance [photon$^2$m$^2$] on the temporally sampled signal. The AO noise error PSD then writes:

\begin{equation}
\label{eq:psd-noise}
    W_\text{res,noise}(\vec{f}) = \frac{\sigma^2_n}{F} \int_{-F/2}^{F/2} \left| \frac{\mathcal{D}(\vec{f}) \, \mathcal{C}(\nu) \, \mathcal{R}(\vec{f})}{1+\Omega(\vec{f},\nu)} \right|^2 d\nu
\end{equation}

\subsubsection{Frozen flow hypothesis}
\label{sec:frozen-flow}

We make the assumption that the atmosphere is the sum of independent turbulent layers, each one sliding with a given wind speed and direction, this is known as the frozen flow hypothesis. Despite the frozen flow hypothesis, considering a high number of layers provides to the simulated atmosphere not only a sliding aspect but also a boiling aspect due to the different wind speeds and directions. However, with no loss of generality and for the sake of simplicity within the following equations, we consider only one turbulent layer with a vectorial speed $\vec{V}$. The results are easily generalized afterwards by summation over the layers. The frozen-flow hypothesis for each layer writes:
\begin{equation}
    \phiatm(\vec{f},\nu) = \phiatm(\vec{f}) \, \delta(\nu-\vec{V}\cdot\vec{f})
\end{equation}
with $\delta$ the Dirac distribution. Thanks to this hypothesis, the spatio-temporal error greatly simplifies:
\begin{equation}
    \begin{split}
        W_\text{res,spatio-temporal}(\vec{f}) & = \int_{-\infty}^{+\infty} \frac{\psd{\phiatm(\vec{f},\nu)}}{|1+\Omega(\vec{f},\nu)|^2} d\nu \\
        & = \frac{\psd{\phiatm(\vec{f})}}{|1+\Omega(\vec{f},\vec{V}\cdot\vec{f})|^2}
    \end{split}
\end{equation}
Consequently, for a given spatial frequency $\vec{f}$, the spatio-temporal PSD depends only on the atmospheric PSD and on the response of the AO loop evaluated at the specific temporal frequency $\nu=\vec{V}\cdot\vec{f}$.\\

Regarding numerical implementation, the consequence of the frozen-flow hypothesis it is that there is no need to manage 3-dimensional (2 spatial and 1 temporal) arrays, but a 1-dimensional array and a 2-dimensional array separately, reducing the computation cost of the spatio-temporal error. The numerical complexity drops from $N_\text{spatial}^2\times N_\text{temporal}$ to $N_\text{spatial}^2 + N_\text{temporal} \simeq N_\text{spatial}^2$ where $N_\text{spatial}$ and $N_\text{temporal}$ are the number of numerical points representing respectively the spatial and temporal frequencies.

\subsubsection{\rfe{Simplification of the aliasing term}}

\rfe{The aliasing term of equation \ref{eq:wfs-phase-error} must be simplified for numerical computations. First of all, let's note that one aliases unseen, and thus uncorrected, spatial frequencies, so that $\phires(\vec{f}+\vec{k}/p,\nu)=\phiatm(\vec{f}+\vec{k}/p,\nu)$ for all aliased terms $\vec{k}\in\mathbb{Z}^2\backslash (0,0)$. The input turbulence model thus serves to compute the WFS aliasing.}\\

\rfe{Second, the sum inside the aliasing equation theoretically runs up to infinity in both spatial directions. However, for numerical computations, the sums is limited to a small number of terms. In our case we choose $\vec{k}\in\{-3,-2,-1,1,2,3\}^2$. Each higher term adds numerical computation time for lower amplitude of aliasing. Indeed the Von-K\'arm\'an PSD evolves as $f^{-11/3}$. Moreover, the sensitivity of Fourier filtering wavefront sensors such as the pyramid WFS or the Zernike WFS exhibit a flat sensitivity at high frequencies \citep{chambouleyron2021phd}, then damped by the "sinc" function of the pixel average, further reducing the amplitude of the aliased frequencies.}

\subsection{\rfe{Tutorial} case of a piecewise constant reconstruction}
\label{sec:piecewise-reconstruction}

Equation \ref{eq:wfs-phase-error}, describing the error budget, simplifies when the phase reconstruction is piecewise constant in the spatial-frequency domain. In other words, there exists a domain $\mathbb{F}\subset\mathbb{R}^2$ and a scalar $\gamma\in\mathbb{R}$ such as, for all $\vec{f}\in\mathbb{F}$:
\begin{equation}
    \mathcal{D}(\vec{f}) \, \mathcal{R}(\vec{f}) \, \sensi_\sqcap(\vec{f}) = \gamma
\end{equation}
There is no constraint on the $\mathbb{F}$ domain shape nor size. For example, it can encompass the full frequency space $\mathbb{F}=\mathbb{R}^2$ or be as small as a unique frequency $\mathbb{F}=\{\vec{f}_0\}$. Within the $\mathbb{F}$ domain, the residual error of equation~\ref{eq:wfs-phase-error} becomes:
\begin{equation}
\label{eq:wfs-phase-error-constant}
    \begin{split}
        \phires(\vec{f},\nu) = & \frac{1}{1+\gamma\,\mathcal{C}(\nu)} \phiatm(\vec{f},\nu) - \frac{\gamma\,\mathcal{C}(\nu)}{1+\gamma\,\mathcal{C}(\nu)} \frac{n(\vec{f})}{\sensi_\sqcap(\vec{f})} \\
        & - \frac{\gamma\,\mathcal{C}(\nu)}{1+\gamma\,\mathcal{C}(\nu)} \sum_{\vec{k}\in\mathbb{Z}^2\backslash (0,0)}  \frac{\sensi_\sqcap(\vec{f}+\vec{k}/p)}{\sensi_\sqcap(\vec{f})} \, \phires(\vec{f}+\vec{k}/p,\nu)
    \end{split}
\end{equation}
where $\sensi_\sqcap$ has been introduced in the denominator, under the assumption that it does not vanish over $\mathbb{F}$. If there exists at least one $\vec{f}\in\mathbb{F}$ such as $\sensi_\sqcap(\vec{f})=0$, then $\gamma=0$ and $\gamma/\sensi_\sqcap$ is undetermined. In such case, one needs to substitute $\gamma/\sensi_\sqcap \to \mathcal{D} \, \mathcal{R}$ to avoid the undetermination issue.\\

The consequence of the piece-wise reconstruction is that spatial and temporal filtering are decoupled, the usual error transfer function (ETF) and noise transfer function (NTF) appear in the equation, both affected by a $\gamma$ factor on the temporal open loop.\\

Even though $\gamma$ can take any value, two peculiar cases can be considered:
\begin{itemize}
    \item case $\gamma = 0$ : there is no DM influence, no WFS sensitivity, or, \textit{a fortiori}, no reconstruction in $\mathbb{F}$,
    \item case $\gamma = 1$ : reconstruction is perfect over $\mathbb{F}$, \textit{i.e.} the reconstructor perfectly inverts the interaction matrix $\sensi_\sqcap\,\mathcal{D}$ in the domain $\mathbb{F}$.
\end{itemize}

\begin{figure}[h]
    \centering
    \includegraphics[width=\linewidth]{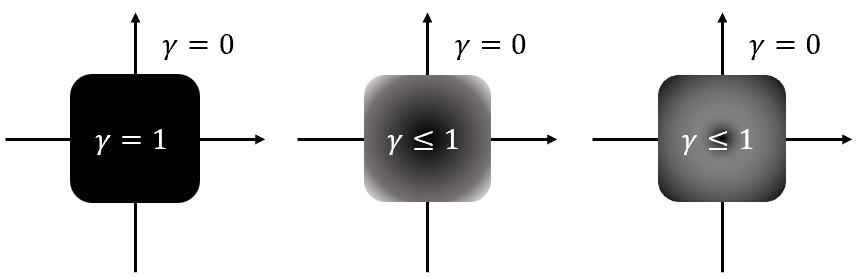}
    \caption{Three different examples of separation of the $(f_x,f_y)$ plane. Left: harsh splitting between the fitting error and the other terms. Middle: smooth transition due to a reconstructor filtering the WFS noise at high spatial frequencies. Right: variations due to optical gains at low and mid frequencies.}
    \label{fig:gamma-cases}
\end{figure}

The area for which $\gamma=0$ produces the so-called fitting error, that is the spatio-temporal error with no temporal contribution. Moreover, noise and aliasing error cancel when $\gamma=0$. The area for which $\gamma>0$ produces all error terms: spatio-temporal, noise and aliasing errors. Enforcing a harsh splitting of the spatial frequency space into two complementary domains $\gamma=0$ and $\gamma=1$ is the technique used in codes like \tiptop (see section~\ref{sec:tiptop}) to split the fitting error from the other error terms. However, any value of $\gamma$ between $0$ and $1$ corresponds to a smooth transition between a no-correction area (fitting error) and an AO corrected area (temporal, aliasing and noise errors). A value $\gamma<1$ can appear intentionally when choosing the optimized modal gain integrator \citep{gendron1994} in order to efficiently filter the noise or for tomographic reconstructors, or may appear unintentionally due to WFS optical gains. Managing optical gains, and thus any value of $\gamma$ is the possibility offered by the present work. Different examples of splitting of the $(f_x,f_y)$ plane are shown on Figure~\ref{fig:gamma-cases}.\\

\subsection{Description of the AO-corrected PSF}

The long-exposure PSF resulting of the AO system correction can be separated into a diffraction static component and a phase residuals component \citep{roddier1981}:
\begin{equation}
    h_\text{AO} = h_\text{diffr} \conv h_\text{res}
\end{equation}
with $h_\text{diffr}$ the diffraction term and $h_\text{res}$ the AO residuals term. This equation holds if the phase statistics are independent of the position in the pupil, thus ignoring pupil edge effects on the turbulence correction, that is usually a valid assumption for AO system \citep{conan1994phd}. The AO component depends on the phase PSD:
\begin{equation}
\label{eq:psd-to-psf}
    h_\text{res} = e^{-\sigma^2_\text{res}} \, \IFFT{e^{\IFFT{W_\text{res}}}}
\end{equation}
with $\IFFT{}$ the inverse Fourier transform, and $\sigma^2_\text{res}=\int W_\text{res}\,d\vec{f}$ the total phase variance. For the numerical implementation, the continuous Fourier transform is replaced by a discrete Fourier transform. The PSF should be sampled at least according to the Shannon-Nyquist theorem to avoid numerical artifacts. Moreover the finite field of view of numerical arrays may ignore energy of fitting error (decrease power law as $f^{-11/3}$) at high frequencies. It must be taken care to encompass enough energy of the fitting error when computing PSD. The equivalent issue for end-to-end simulations is that the limited resolution in the pupil plane cannot describe accurately the high spatial frequencies.\\

Regarding the scientific instrument branch, it suffers from the AO residuals already listed (spatio-temporal, aliasing, noise), but also from extra error terms, such as anisoplanetism between the scientific target and the AO guide star or chromatic effects (differential atmospheric diffraction and chromatic dependence of the refraction index of the atmosphere) if the WFS and science are performed at different wavelengths. All these terms do not impact the AO loop and can be processed afterwards. We group them under the name $W_\text{extra}$. The residual PSD at the scientific target is then given by:
\begin{equation}
    W_\text{science}(\vec{f}) = W_\text{res}(\vec{f}) + W_\text{extra}(\vec{f})
\end{equation}
Using this PSD within equation~\ref{eq:psd-to-psf} allows to compute the science PSF. Moreover the static component $h_\text{diffr}$ in the scientific branch may include non common path aberrations (NCPA) if they are estimated or measured \citep{ndiaye2016} for a given instrument. The differences between the AO branch and scientific branch are shown on Figure~\ref{fig:method-og}.

\subsection{FF-WFS sensitivity and non-linearity}

A wavefront sensor is non-linear if its sensitivity evolves with the phase to be measured. Within the linear parameter-varying system (LPVS) formalism, the measurement is not anymore $\sensi\times\varphi$ but $\sensi_\varphi\times\varphi$, with $\sensi_\varphi$ the sensitivity of the WFS towards the current phase $\phi$. Fourier-filtering wavefront sensors (FF-WFS) such as the pyramid WFS, the Zernike WFS or the Bi-O edge WFS are highly sensitive and prone to such non-linear behaviour. The LPVS formalism can be applied to FF-WFS \citep{fauvarque2017phd, chambouleyron2021phd} and lead to the so-called optical gains. The latter are defined as the ratio between current and calibration sensitivities $\text{OG}=\sensi_\sqcap/\sensi_\sqcap^\text{calib}$. The current sensitivity $\sensi$ of the WFS can be evaluated through the PSF seen by the WFS \citep{fauvarque2019, chambouleyron2021}. Let's first define the impulse response of the FF-WFS:
\begin{equation}
\label{eq:impulse-response}
    H_\text{AO} = 2\, \text{Im}\left\{ \overline{\widehat{m}} \, (\widehat{m} \conv \widehat{\,\omega\conv h_\text{AO}\,}) \right\}
\end{equation}
with $h_\text{AO}$ the current PSF, $\omega$ the modulation path and $m$ the complex mask of the FF-WFS. According to \cite{fauvarque2019}, the sensitivity then writes:
\begin{equation}
\label{eq:sensitivity}
    \sensi(\vec{f}) = \alpha_n \sqrt{\left|\widehat{H_\text{AO}}\right|^2 \conv h_\text{diffr}}
\end{equation}
with $h_\text{diffr}$ the diffraction limit PSF and $\alpha_n$ the noise amplification factor, that equals one for readout noise and the square root of the number of faces for a modulated pyramid \citep{chambouleyron2021phd}. This equation thus provides a method to compute the sensitivity of the FF-WFS for any PSF ($h_\text{AO}$) seen by the WFS. Estimation of the sensitivity thanks to the PSF corresponds to the Gain Scheduling Camera technique \citep{chambouleyron2021}. Even though this method relies on the hypothesis of the convolutive formalism, in particular it ignores possible modal confusion, it has been shown accurate for both simulations and on-sky experiments \citep{striffling2025}.\\

\begin{figure}[h]
    \centering
    \includegraphics[width=\linewidth]{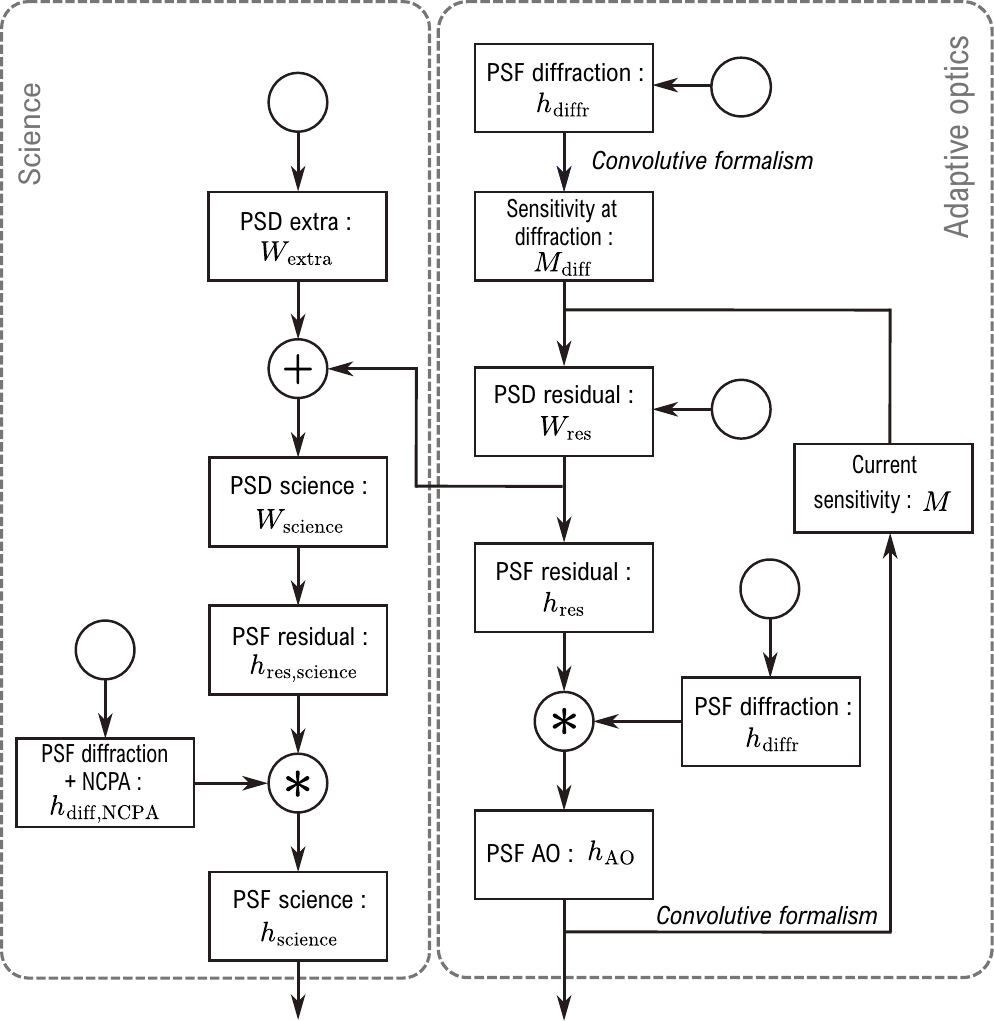}
    \caption{Iterative method to compute the FF-WFS sensitivity, making the modelled FF-WFS non-linear. The AO residuals estimation (right part) is run before the science estimation (left part). Scenario inputs (turbulence, AO system, guide object) required to initialize the method are represented by empty circles.}
    \label{fig:method-og}
\end{figure}

The issue arising for a numerical implementation is that the sensitivity depends on the current PSF (equation~\ref{eq:sensitivity}) but the current PSF depends on the sensitivity (equations~\ref{eq:psd-noise} and \ref{eq:psd-to-psf}), thus creating a feedback loop between sensitivity and PSF. This is solved thanks to the following method \citep{fetick2023}:
\begin{enumerate}
    \item Initialization: assume a diffraction limit PSF on the WFS ($h\equiv h_\text{diffr}$) to compute the WFS sensitivity $\sensi_\text{diffr}$ (equations \ref{eq:impulse-response} and \ref{eq:sensitivity})
    \item Compute the residual PSD for the given sensitivity
    \item Compute the (long-exposure) PSF seen by the WFS from the residual PSD (equation \ref{eq:psd-to-psf})
    \item Update the estimation of the sensitivity given the PSF on the WFS (equations \ref{eq:impulse-response} and \ref{eq:sensitivity})
    \item Iterate on steps 2, 3 and 4
\end{enumerate}
These steps are shown on Figure~\ref{fig:method-og}. In practice the method converges in approximately 2 iterations on the sensitivity estimation (see appendix~\ref{sec:convergence}), making it fast and adapted to numerical implementation. The algorithm is setup by default with 3 iterations.

\subsection{Numerical implementation}
\label{sec:numerical-implementation}

The equations of this section and the management of non-linearities have been implemented into a Python package called \aopera. Both the general 3D form of equation~\ref{eq:wfs-phase-error} and the simplified 2D+1D form of equation~\ref{eq:wfs-phase-error-constant} have been numerically implemented to provide the choice between a more accurate or faster computation. The \aopera package is available on GITLAB\footnote{\href{https://gitlab.lam.fr/lam-grd-public/aopera}{https://gitlab.lam.fr/lam-grd-public/aopera}} for continuous developments or issues tracking, and on PIPY\footnote{\href{https://pypi.org/project/aopera/}{https://pypi.org/project/aopera/}} (\texttt{pip install aopera}) for public releases.

\section{Positioning with respect to Fourier-based AO PSF tools}
\label{sec:tiptop}

Among Fourier-based analytical tools for AO PSF computation, \tiptop \citep{neichel2021tiptop} has emerged as a reference framework thanks to its versatility, computational efficiency, and adoption within several VLT instrument studies such as ERIS and MAVIS. It is currently being adopted by multiple facilities and instrument teams in preparation for the ELT, and is used as a baseline tool for HARMONI MCAO mode performance estimation \citep{thatte2021, harmoni2026_mcao}.\\

Within the \textit{astro-tiptop}\footnote{https://github.com/astro-tiptop/TIPTOP} ecosystem, \tiptop acts as a high-level interface relying on lower-level PSD-based modelling packages such as P3 and, more recently, TipTorch \citep{kuznetsov2026}. In this context, \aopera, being compatible with this framework in SCAO mode, can be considered as a complementary modelling component.\\

While \tiptop supports pyramid WFS (PWFS) configurations, non-linear wavefront sensing effects are not explicitly included in its current analytical applicability domain. In particular, the PWFS is described through an analytical Fourier-domain filter, consistent with classical PSD-based formulations \citep{VERINAUD200427, Correia:17}, where the sensor sensitivity is assumed fixed.\\ 
As described above, the approach proposed in this paper instead relies on a convolutive sensitivity formalism combined with an optical gain feedback loop, and is expected to enable more accurate AO performance estimations.\\ This makes \aopera a suitable candidate for integration within the \textit{astro-tiptop} ecosystem, which is currently under discussion, especially given the role of \tiptop as a reference tool for ELT applications and for instruments such as HARMONI operating in PWFS-based SCAO modes \citep{harmoni2026_scao}.\\

In addition, recent developments enable \aopera to run the most computationally intensive parts efficiently on either CPU or GPU (NVIDIA CUDA-enabled).\\

\section{Validation and comparative results}
\label{sec:results}

\subsection{Test framework}

The \oopao\footnote{https://github.com/cheritier/OOPAO} software (Object-Oriented  Python  Adaptive  Optics) \citep{heritier2023oopao}, inheriting from \texttt{OOMAO} \citep{conan2014oomao}, is a end-to-end simulation tool. The software draws random phase screens and moves them across the telescope pupil, the wavefront is compensated by the influence functions of the deformable mirror and the residuals are propagated to a WFS and to its camera. On the camera, different noises can be added, such as the photon shot noise or readout noise. At each time step, at least as fast as the AO loop frequency, the phase screens are moved, the residuals propagated to the WFS, commands computed and sent to the DM. This method makes the tool quite close to the actual propagation of light through an AO system. The methodology and numerical implementation of \oopao being completely different from the ones of this paper, it makes \oopao suitable for comparative tests. The \oopao simulation is set-up with the parameters defined in Table~\ref{tab:simulation-param} representing the baseline test case. The condition parameters are then modified in the next sections with respect to this baseline. It allows to explore different operation cases and to validate the method over a wide range of scenarios. 

\begin{table}[h]
    \centering
    \begin{tabular}{lcc}
        \hline\hline
        Parameter & Unit & Value \\
        \hline
        Telescope diameter & m & $1.5$\\
        Nb. actuators across diameter & $-$ & 21\\
        AO loop frequency & Hz & $1000$\\
        Integrator gain & $-$ & $0.4$\\
        Frame delay & $-$ & 1\\
        NGS band & $-$ & $I \, (790\text{ nm})$\\
        Modulation radius & $\lambda/D$ & 3\\
        Fried parameter at 500 nm & cm & $5$ or $7$\\
        Number of layers & $-$ & 3\\
        Wind speed & m/s & $[8.7, 10.4, 13]$\\
        Layer fractional $C_n^2$ & $-$ & $[0.6, 0.2, 0.2]$\\
        OOPAO nb. pixels across diameter & $-$ & $168$ \\
        \hline
    \end{tabular}
    \caption{Baseline parameters for \oopao and \aopera simulations.}
    \label{tab:simulation-param}
\end{table}

\subsection{Performance versus wind speed}

At high SNR on the wavefront sensor, the error budget is limited by the spatio-temporal and aliasing errors. Variation of the wind speed balances the fitting and aliasing terms with respect to the temporal term. The \oopao simulation is run with a guide star magnitude $m_I=0$, corresponding to a high SNR case.\\

\begin{figure}[h]
    \centering
    \includegraphics{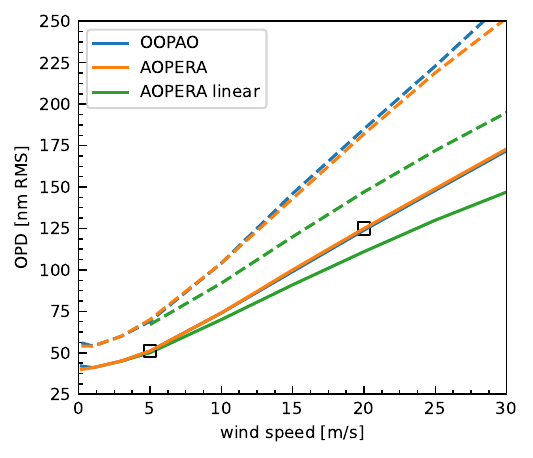}
    \caption{Optical path difference OPD [nm] versus the equivalent wind speed for \oopao and \aopera. The guide star magnitude is set to $m_I=0$ corresponding to a high SNR case. Two cases of $r_0=5$ cm (dashed lines) and $r_0=7$ cm (plain lines) are considered. Black squares correspond to the PSF seen on Figure~\ref{fig:psf-aopera-oopao}.}
    \label{fig:opd-vs-windspeed}
\end{figure}

Let's define the equivalent wind speed of the turbulence:
\begin{equation}
    \overline{V} = \left( \frac{\int C_n^2(h)v^{5/3}(h)\,dh}{\int C_n^2(h)\,dh} \right)^{3/5}
\end{equation}
The wind profile of Table~\ref{tab:simulation-param} is scaled to obtain a desired equivalent wind speed, and one simulation is launched for each equivalent wind speed. Results of \oopao and our method versus the equivalent wind speed are shown on Figure~\ref{fig:opd-vs-windspeed}. Our method is validated with respect to the end-to-end simulation over a wide range of wind speeds and performance.\\

Moreover, the WFS can be artificially made linear by keeping the diffraction limit sensitivity for the WFS instead of updating it (cutting the feedback loop in Figure~\ref{fig:method-og}). Doing so, the spatio-temporal error is not impacted by optical gains, and the AO error budget is under-estimated (green curves on Figure~\ref{fig:opd-vs-windspeed}). This leads to the dramatic relative error of $30\%$ on the RMS OPD, and $56\%$ on the variance, in the worst case of $\overline{V}=30$ m/s and $r_0=5$ cm. The relative error on the variance is still as high as $24\%$ for $r_0=5$ cm and moderate wind speed of $\overline{V}=10$ m/s. Consequently, there is a necessity to include the optical gains in the PSD based models, especially for WFS working in lower Strehl regime.\\

\begin{figure}[h]
    \centering
    \includegraphics{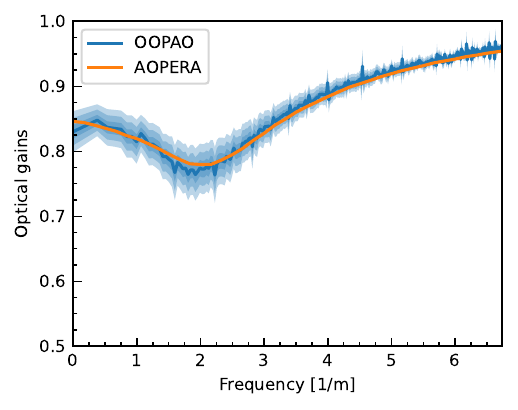}
    \caption{Optical gains computed with \oopao and \aopera in the case $r_0=7$ cm and $\overline{V}=20$ m/s. Shaded blue area show $\pm 1$ to $3$ std of \oopao optical gains across PSF short exposure frames.}
    \label{fig:og}
\end{figure}

Computation of optical gains on short exposure frames with \oopao and long exposure frame with \aopera is shown in Figure~\ref{fig:og}. The optical gains are well estimated by \aopera resulting in a correct assessment of the loss of sensitivity impacting the temporal error (factor $\gamma$ in section~\ref{sec:piecewise-reconstruction}), leading to a good estimation of the global AO performance.\\

The PSF computed with \aopera and \oopao are shown on Figure~\ref{fig:psf-aopera-oopao} in the case of low and high wind speed (moreover, see appendix~\ref{sec:psd_and_psf} for PSD and PSF side views). The shape of the PSF and PSD computed with the two methods are matching in different regimes: fitting error dominated at low wind speed or temporal error dominated at high wind speed. The similarity of shape with end-to-end simulations includes the diffraction pattern, the correction area, the fitting error, the aliasing (visible as a faint cross inside the correction area at low wind speed), or the PSF elongation due to high wind speed. The resulting PSF from \aopera are thus accurate and can be used for AO analysis or scientific data simulation and analysis: exposure time calculator, generation of mock-up scientific data, or post-processing.

\begin{figure}[h]
    \centering
    \includegraphics[width=\linewidth]{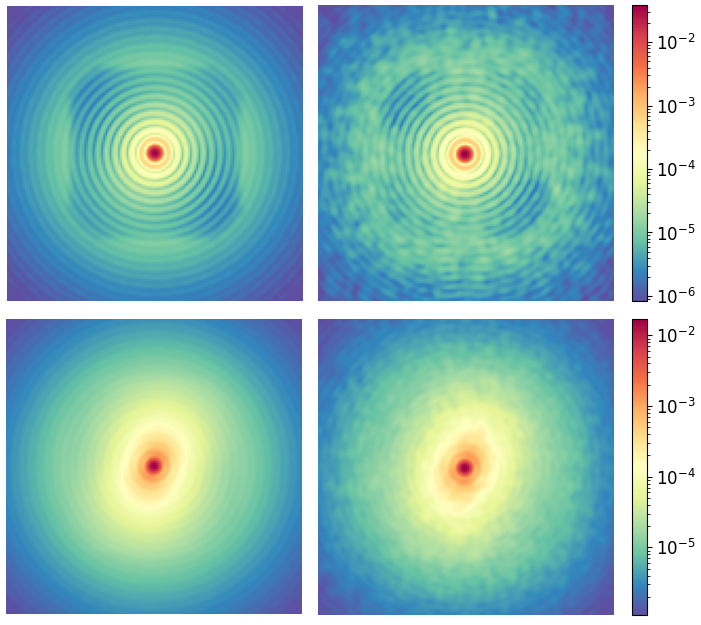}
    \caption{Long exposure PSF at $790$ nm, computed with \aopera (left) and \oopao (right) for a magnitude $m_I=0$ and $r_0=7$ cm. Top: $\overline{V}=5$ m/s. Bottom: $\overline{V}=20$ m/s.}
    \label{fig:psf-aopera-oopao}
\end{figure}

\subsection{Performance versus magnitude}

\begin{figure}[h]
    \centering
    \includegraphics{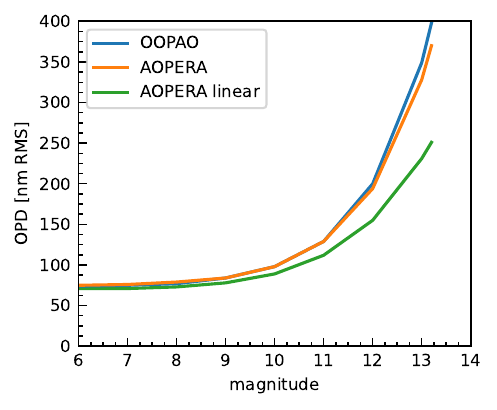}
    \caption{Optical path difference OPD [nm] versus the guide star magnitude for \oopao and \aopera. The equivalent wind speed is set to $\overline{V}=10$ m/s and $r_0=7$ cm.}
    \label{fig:opd-vs-magnitude}
\end{figure}

The performance of the WFS versus the guide star magnitude is critical to assess the limit magnitude of the AO system, defining its sky coverage. This is consequently a critical metric for any AO design. Performance versus magnitude also allows to choose the correct guide stars for AO operations and for exposure time calculators (ETC). A precise estimation of the WFS sensitivity is required, this is provided here by the convolutive formalism. Management of optical gains is of importance since they reduce the sensitivity of FF-WFS and thus the sky coverage of the AO system. The performance of the AO system versus the sensitivity is shown on Figure~\ref{fig:opd-vs-magnitude}. The end-to-end simulations and our PSD based method give similar results. The position of the inflection point versus the magnitude is of great importance since it reveals a modification in the operating regime of the WFS and sets the limit magnitude of the overall AO system. The linear approximation (green curve on Figure~\ref{fig:opd-vs-magnitude}) leads to an erroneous estimation of the limit magnitude, eventually over-estimating the sky coverage of the instrument.

\section{Applications}
\label{sec:applications}

\subsection{Optimal modulation abacus}

A major advantage of the modulated pyramid WFS is that the sensitivity versus linearity trade-off can be tuned thanks to the modulation radius. The optimal modulation radius evolves depending on the magnitude of the guide star and the turbulence conditions.\\

\begin{figure}[h]
    \centering
    \includegraphics{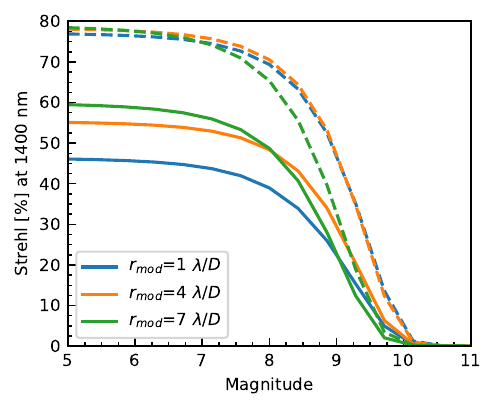}
    \caption{Strehl ratio versus magnitude for different modulation radii for the PAPYRUS AO system. Plain lines: seeing of $2$ arcsec. Dashed lines: seeing of $1$ arcsec.}
    \label{fig:modulation}
\end{figure}

We consider the case of the PAPYRUS adaptive optics system \citep{muslimov2021,fetick2023papyrus} installed at Observatoire de Haute Provence (OHP, France). The AO system is made of a deformable mirror $17\times 17$, a visible modulated pyramid, and a real-time controller running at $500$ Hz. Figure \ref{fig:modulation} shows the predicted Strehl ratio of the PAPYRUS adaptive optics instrument with respect to the star magnitude, for two different seeing conditions. In case of good seeing, the best Strehl ratio is obtained for low modulation radii for all magnitude $M>7$, since the system is linear and consequently requires high sensitivity. However the conclusion is different in case of stronger turbulence (bottom graph). Indeed, at high flux, the system gives better performance at higher modulation radius, favoring the linearity. For lower flux, the modulation radius should be decreased to improve the sensitivity. The management of linearity and sensitivity is critical for such applications, where the conclusion is not straightforward and the performance should be numerically computed for each AO system and observing conditions. A non-linear tool such as AOPERA is necessary here, since a linear tool would always favour the most sensitive WFS, thus requiring the lowest modulation radius, and leading to erroneous conclusions regarding instrumental setup. The benefit of a non-linear PSD based method is to be able to compute the optimum of an AO system on the fly, during on-sky observations, following the evolution of the turbulence during the night. Such a method is suited for a numerical twin of the AO system to be run in parallel of observations, with a low computational cost.\\

Comparison between PAPYRUS experimental data and simulation tools is available in \cite{alagao2026}. Processing of experimental data is out of the scope of the current paper since it relies on external tools such as seeing monitor and wind speed estimators, so that comparison with simulation tools is not straight forward and might suffer from estimators' biases.

\subsection{Extended sources observation}

Observation of extended sources within the Solar System is envisaged for ELT instruments \citep{delsanti2023}. Moreover, the PROVIDENCE telescope \citep{petit2024} will be built in France with time slots dedicated to resolved observation of Low-Earth Orbit (LEO) satellites. There is consequently a need to quantify the performances of such AO systems when looking at extended objects. AO design studies or exposure time calculators must accurately model the response of FF-WFS on extended objects. The pyramid WFS suffers from a dramatic loss of sensitivity whenever looking at extended sources \citep{iglesias2002,pinna2011,ellerbroek2017,oyarzun2024}. Indeed, the spatial extension of the source is a collection of incoherent points similar to a modulation pattern for the P-WFS. It thus falls within the convolutive formalism of \cite{fauvarque2019}. The weight function $\omega$ defined in the FF-WFS impulse response (equation~\ref{eq:impulse-response}) not only includes the modulation path but also the shape of the AO guide object. The main limitation of this formulation is that it does not include possible anisoplanetism on the WFS measurement, and it is thus limited to object' size smaller than the isoplanetic angle.\\

\begin{figure}[h]
    \centering
    \includegraphics{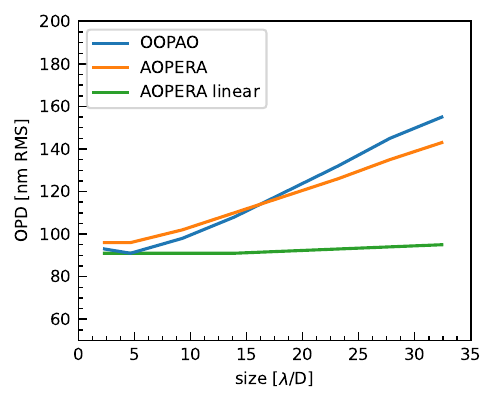}
    \caption{Optical path difference OPD [nm] versus the source diameter. The modulation radius is $3\lambda/D$ at calibration, and turned off for on-sky observations.}
    \label{fig:opd-vs-size}
\end{figure}

Source extension is taken into account in simulations with the following procedure: the AO system is calibrated with a circular modulation of radius $3\lambda/D$, the modulation is inactivated on-sky and replaced by a disk like source. For \oopao, the disk is a collection of points separated by $0.5\lambda/D$ on X and Y axis for accurate dense description of the source. Bigger objects respectively to $\lambda/D$ thus require more computation with end-to-end simulations, whereas the numerical complexity is constant for our \aopera method based on convolutional formalism. Results are shown on Figure~\ref{fig:opd-vs-size}. If the optical gains are not taken into account, the performance does not depend on the source size, indeed the WFS is supposed linear, and thus neither the spatio-temporal error nor the noise error depend on the WFS supposedly constant sensitivity. It is thus necessary to take into account the optical gains to describe the effect of extended objects on the FF-WFS performance. Both codes \oopao and \aopera have the same tendency showing that increasing the object size decrease the performance, however \aopera is slightly less sensitive to the source extension, and is valid only for moderate variations of the source size with respect to the calibration.

\subsection{Optical gain compensation}

\begin{figure}[h]
    \centering
    \includegraphics{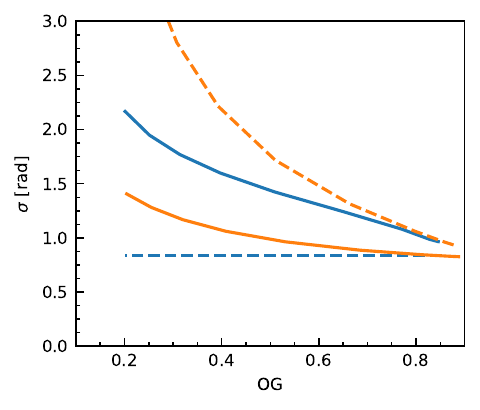}
    \caption{Temporal error (blue) and noise error (orange) versus optical gains for a PAPYRUS like case. Two control strategies are considered: no compensation of optical gains (plain lines) and compensation of optical gains (dashed lines). Seeing and star magnitude are chosen so that temporal and noise errors match at high optical gain values for easier relative comparison of the curves.}
    \label{fig:error-vs-og}
\end{figure}

Due to the WFS non-linearity, optical gains may appear. They are cause by a mismatch between the reconstructor, computed for a given sensitivity at calibration $\sensi_\sqcap^\text{calib}$, and the on-sky sensitivity of the WFS. In most cases, optical gains are below unity, and the reconstructed phase is under-estimated, leading to $\gamma<1$ (with the notation of section~\ref{sec:piecewise-reconstruction}). Multiple techniques exist to estimate the optical gains during AO loop operations \citep{deo2019,esposito2020,chambouleyron2021}. With such techniques, the reconstructor can be updated to compensate for optical gains during AO loop operations, and eventually restore $\gamma=1$. Even though the spatio-temporal error has been lowered to its original value (as if there where no optical gains) by the optical gain compensation, the noise error dramatically increases, as seen on figure~\ref{fig:error-vs-og}. Indeed, the sensitivity has been reduced, and the reconstructor increased to compensate for the optical gains, further increasing the noise error. Once again, this effect can be modelled with non-linear tools only.

\section{Conclusions}

In this article, we have developed a formalism to accurately estimate the AO residual PSD and AO corrected PSF. The convolutive formalism for Fourier-filtering WFS allows for a precise estimation of their sensitivity. Moreover, the sensitivity is included into a numerical feedback loop to manage the optical gains effects and to produce a non-linear wavefront sensor within a PSD based approach. The method has been validated against end-to-end simulations to show its fidelity over a wide range of observing conditions.\\

The method and the tool can be used to describe the performance of future AO systems including Fourier filtering WFS, such as the pyramid WFS or the Zernike WFS. These WFS, due to their exquisite sensitivity, will be extensively used for ELT instruments. Moreover, they are particularly adapted to extreme AO and double stage AO systems, with a scientific goal of high contrast or single mode fiber injection to produce high resolution spectra. The technique developed through this article may be used for such extreme AO applications.


\begin{acknowledgements}
  This work benefited from the support the French National Research Agency (ANR) with the Programme Investissement Avenir F-CELT (ANR-21-ESRE-0008), the ANR-DGA-AID ASTRID program AI4AO  (ANR-25-ASTR-0015), the Action Spécifique Haute Résolution Angulaire (ASHRA) of CNRS/INSU co-funded by CNES, the french government under the France 2030 investment plan (cassiopée project) and the Initiative d’Excellence d’Aix-Marseille Université A*MIDEX, program number AMX-22-RE-AB-151. The PROVIDENCE project is supported by the European Regional Development Fund (ERDF), under the Provence-Alpes-Côte d’Azur Region and Alps Massif / ERDF-ESF+JTF 2021-2027 program, to support projects contributing to the economic, scientific, and technological development of the territory. Authors credit and thank Lucie Jonkisz for the AO block-diagram illustrations.
\end{acknowledgements}

\bibliographystyle{template/aa}
\bibliography{biblio.bib}

\begin{appendix}

\section{Convergence of the optical gains estimation}
\label{sec:convergence}

\begin{figure}[h]
    \centering
    \includegraphics{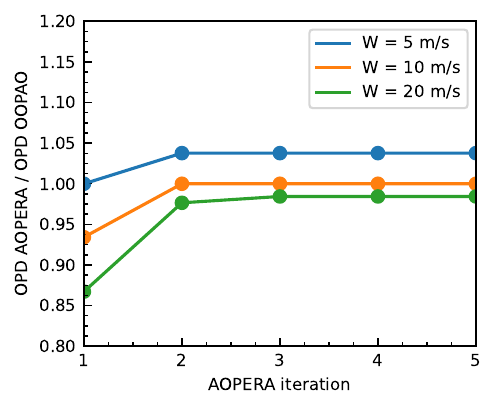}
    \caption{Convergence of the OPD versus the number of optical gain iterations. Simulations are run with parameters of Table~\ref{tab:simulation-param}.}
    \label{fig:convergence}
\end{figure}

The loop on the optical gains estimation with the AOPERA method converges in two iterations only, as seen on the Figure~\ref{fig:convergence}. Results are shown for three different values of wind speed, corresponding to good to low performance regime (Figure~\ref{fig:opd-vs-windspeed}). The algorithm runs by default over three iterations as a compromise between accuracy and speed. The relative error between the OPD computed after three iterations and its final value (five iterations on the graph) is less than $1\%$.

\section{PSD and PSF side views}
\label{sec:psd_and_psf}

\begin{figure}[h]
    \centering
    \includegraphics{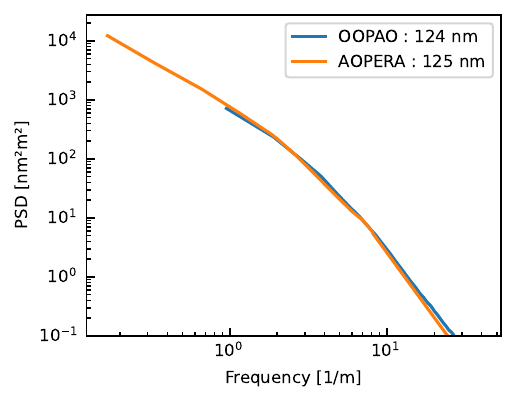}
    \includegraphics{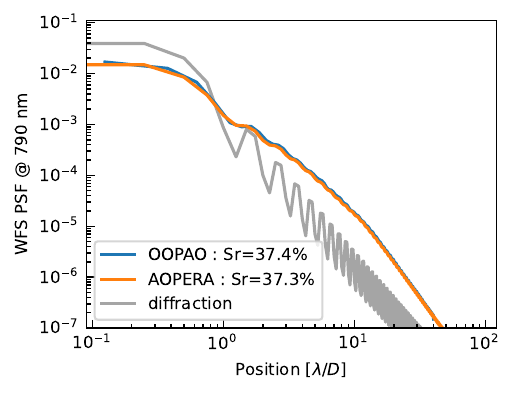}
    \includegraphics{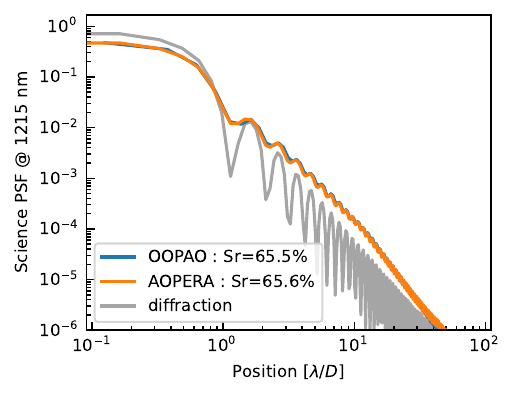}
    \caption{From top to bottom: PSD, WFS PSF and scientific PSF. The wind speed is set to $\overline{V}=20$ m/s, $r_0=7$ cm and $m_I=0$.}
    \label{fig:psd-psf}
\end{figure}

The Strehl ratio is a simple metric to quantify the performance of the AO system, however it does not state about the distribution of energy in the focal plane. The PSD of the phase is important for high contrast and coronagraphic applications. Figure~\ref{fig:psd-psf} shows the PSF and PSD computed with the \aopera and \oopao codes. The two codes produce a similar PSD. The resolution in the frequency domain is different though. Indeed for \aopera the frequency step is defined from the required PSF sampling and the diameter such as $\delta f=1/(D\,s)$ with $D$ the telescope diameter and $s$ the required PSF sampling. However, for \oopao, the PSD is computed from the residual phases projected onto the pupil. In order to remove the pupil edge effects within end-to-end simulations, the Fourier transform of the phase screens are computed within the square of diagonal $D$ inscribed in the circular pupil of diameter $D$. This method ignores a part of edge effects and assumes that the PSD is homogeneous in the pupil. This is the case for our simulations since the PSF, the Strehl ratio and the phase standard deviation are matching when computed on the full circular pupil.

\end{appendix}

\end{document}